\documentstyle[prd,aps,psfig]{revtex}

\def\simlt{\stackrel{<}{{}_\sim}}  
\def\bma#1{\mbox{\boldmath{$#1$}}}

\newcommand\be{\begin{equation}}  \newcommand\ee{\end{equation}}
\newcommand\bea{\begin{eqnarray}}  \newcommand\eea{\end{eqnarray}}
\newcommand\ba{\begin{array}}  \newcommand\ea{\end{array}}
\begin{document}
\draft \input epsf  \def\la{\mathrel{\mathpalette\fun <}}
\def\ga{\mathrel{\mathpalette\fun >}}
\def\fun#1#2{\lower3.6st\vbox{\baselineskip0st\lineskip.9st
\ialign{$\mathsurround=0pt#1\hfill##\hfil$\crcr#2\crcr\sim\crcr}}}

\twocolumn[\hsize\textwidth\columnwidth\hsize\csname
@twocolumnfalse\endcsname

\title{New supersymmetric source of neutrino masses and mixings}
\author{J.A. Casas$^{(1, 3)}$, J.R. Espinosa$^{(2,3)}$  and
I. Navarro$^{(1)}$} \address{\phantom{ll}} \address{$^{(1)}${\it IEM,
CSIC, Serrano 123, 28006 Madrid, Spain}} \address{$^{(2)}${\it IMAFF,
CSIC, Serrano 113 bis, 28006 Madrid, Spain}} \address{$^{(3)}${\it
IFT, C-XVI, Univ. Aut\'onoma de Madrid, 28049 Madrid, Spain}}
\date{\today}  \maketitle
\begin{abstract}
Conventionally, neutrino masses in a supersymmetric theory arise from
non-renormalizable lepton-number (L)-violating operators in the 
superpotential. The
alternative possibility of having such operators in the K\"ahler
potential as the dominant source of neutrino masses has very
interesting implications and differences with respect to the standard
scenario: first, the scale of L-violation can be lowered dramatically
and neutrino masses  have a different $\tan\beta$-dependence; second,
the renormalization of these operators has remarkable properties: 
in many cases it improves drastically the stability of neutrino textures 
against radiative corrections, while in others it makes possible to 
generate radiatively large mixing angles in a natural way. In particular, 
the mass splitting and mixing angle of solar neutrinos (LAMSW solution) 
can be explained as a purely radiative effect.
\end{abstract}
\pacs{PACS: 14.60.Pq, 12.60.Jv~ \hskip3.5pc IFT-UAM/CSIC-02-21,~
IEM-FT-226/02.  [hep-ph/0206276]} ]


{\bf 1. Introduction.}  There is by now compelling evidence of
neutrino oscillations in atmospheric and solar fluxes
\cite{SNO,SK}, which gives strong evidence for neutrino masses with
substantial mixing.  The mass splittings and mixing angles required by
observations are, at $3\sigma$ \cite{concha}   $2.4\times
10^{-5}<\Delta m_{sol}^2/(eV)^2 <2.4\times 10^{-4}$,   $0.27<
\tan^2 \theta_{sol}< 0.77$ (Large-Angle MSW explanation  of the solar
fluxes, as favoured by the SNO data \cite{SNO}); and $1.4\times
10^{-3}<\Delta m_{atm}^2/(eV)^2<6.0 \times 10^{-3}$,   $0.4<\tan^2
\theta_{atm}< 3.0$.    Neutrino masses can be Dirac or Majorana, but
the latter case is more appealing, as it provides an explanation for
the smallness of $m_\nu$. In the Standard Model (SM) the lowest
operator describing Majorana $\nu$-masses is $(L\cdot H\ L\cdot H)/M$,
where $L^T=(\nu, e)_L$ (flavor indices suppressed), $H$ is the Higgs 
doublet and  $M$ is the scale of new physics 
(that violates lepton number, L).  For $M\sim 10^{14}$
GeV, (not far from the GUT scale),   $m_\nu = \langle H \rangle^2 / M$
is of the right order of magnitude.

This scheme, nicely realized by the see-saw mechanism,  can be easily
made supersymmetric (SUSY). This is desirable,  as the presence of
large scales stresses the SM hierarchy problem. The standard SUSY framework
has an operator $(L\cdot H_2)^2/M$ in the
superpotential $W$, giving $m_\nu= \langle H_2 \rangle^2 / M=v^2\sin^2
\beta/(2M)$ (with $v=246$ GeV, $\tan \beta=\langle H_2 \rangle/\langle
H_1 \rangle$).

In this letter we explore an alternative source of $\nu$-masses not
usually considered: $\nu$-masses arising from operators in the
K\"ahler potential, $K$. We will enumerate the main results from 1)
to 7).

{\bf 2. Neutrino masses from ${\bma K}$.}  In the K\"ahler potential,
the non-renormalizable operators of lowest dimension that violate
L-number have dimension 4 and are
\bea
\label{KL}
K&\supset& {1\over 2M^2}\kappa_{ij}(L_i\cdot H_2)(L_j\cdot
\overline{H}_1)\nonumber\\ &+&{1\over 4M^2}\kappa'_{ij}(L_i\cdot
\overline{H}_1)   (L_j\cdot \overline{H}_1)+ {\mathrm h.c.}\ , \eea
where $M$ is the scale of L-violating new physics  and $\kappa$,
$\kappa'$ are adimensional matrices in flavour space.  We do not
include in $K$ the bilinear operators $(L_i\cdot \overline{H}_1)$
which break $R$-parity (and L-number) already at the  renormalizable
level. Dimension-4 operators of the form  $(L_i\cdot H_2)(L_j\cdot
H_2)$ do not contribute to the Lagrangian (being holomorphic) and can
be ignored (in global SUSY).  Finally, an operator
$\kappa''_{ij}(L_i\cdot L_j)(H_2\cdot \overline{H}_1)/M^2+ {\mathrm
h.c.}$ in $K$ can always be absorbed in the  antisymmetric part of the
operator $\kappa$ due to the identity $\kappa''_{ij}(L_i\cdot
L_j)(H_2\cdot \overline{H}_1)= (\kappa''_{ij}-\kappa''_{ji})(L_i\cdot
H_2)(L_j\cdot \overline{H}_1)$.  Notice that, while $\kappa'$ is a
symmetric matrix, $\kappa$ may contain a symmetric and an
antisymmetric part: $\kappa\equiv  \kappa_S+\kappa_A$.

These operators in the K\"ahler potential generate neutrino masses
after electroweak symmetry breaking, whenever $\langle F^*_{H_1}
\rangle\sim -\partial W /\partial H_1\neq 0$, which indeed occurs due
to the $\mu H_1 H_2$ term in  $W$.  The relevant term in the
Lagrangian reads:
\be \delta {\cal L}={1\over 2}{\partial^3 K\over \partial\phi_I
\partial\phi_J \partial\phi_L^*}{\cal G}^{-1}_{LK}{\partial
W\over\partial\phi_K}(\chi_I\cdot\chi_J) +{\mathrm h.c.}\ ,
\label{LL} 
\ee
where the Weyl spinors $\chi_I$ are the fermionic partners of the
scalar fields $\phi_I$; $(\chi_I\cdot\chi_J)\equiv
i\chi_I^T\sigma_2\chi_J$; ${\cal G}_{IJ}\equiv\partial^2
K/\partial\phi_I\partial\phi_J^*$.  Plugging (\ref{KL}) in (\ref{LL})
one obtains Majorana mass terms for neutrinos,  ${1\over
2}{M_\nu}_{ij}(\nu_i\cdot\nu_j)+{\mathrm h.c.}$, with
\be M_\nu={\mu v^2\over M^2} \left[ \kappa_S\sin^2\beta
+\kappa'\sin\beta\cos\beta\right]  \ .
\label{nufromK}
\ee  Notice that only the symmetric part of $\kappa$, {\it
i.e.}~$\kappa_S$, contributes to the neutrino mass matrix, while
$\kappa_A$ does not (in analogy with the mechanism of
ref.~\cite{Arcadi}).
 
Eq.~(\ref{nufromK}) shows two important differences of SUSY  Majorana
neutrino masses coming from the K\"ahler potential with respect to
the usual mechanism ($M_\nu$ from $W$):
\begin{description}

\item [ 1)] The dependence on $\tan\beta$ is different from that of
the conventional MSSM [$m_\nu(\beta)\sim \sin^2\beta$].

\end{description}

\noindent The new term $\kappa'\sin\beta\cos\beta$ goes to zero for
large $\tan\beta$ so that, if only this operator were present, one
could get an extra suppression in neutrino masses.

\begin{description}
\item [ 2)] There is an additional ${\cal O}(\mu/M)$ suppression 
in $m_\nu$ with respect to $\nu$-masses from $W$ ($m_\nu\sim v^2/M$).

\end{description}

\noindent The fact that $\mu\simlt {\cal O}(TeV)$ (for  the
naturalness of electroweak breaking) implies that one can now lower
significantly the scale  $M$ of L-violation.  Instead of
the typical range   $M\sim 10^{13}-10^{15}$ GeV (which gives
 $m_\nu\sim 1-10^{-3}$ eV) one can easily get $M\sim
10^{7}-10^{9}$ GeV.  These figures assume that all adimensional
$\kappa$'s in (\ref{nufromK}) are of order 1. If they are further
suppressed, {\it e.g.}~if they are generated at the loop level or
through small couplings (maybe of non-perturbative origin)  or are
affected by the $\tan\beta$ suppression described above,  then the
scale $M$ can be much closer to the electroweak scale.  This is very
interesting for experimental reasons and to make contact with recent
theoretical scenarios which favor such relatively light scales of new
physics: these scenarios may accommodate realistic
neutrino Majorana masses in a natural way, provided the relevant
operators live in $K$ instead of~$W$.

{\bf 3. ${\bma K}$ versus ${\bma W}$.}  There can be at least two
reasons for neutrino  masses to originate in $K$  rather than in 
$W$.  (As we have seen, if both sources were present
and were suppressed by  the same mass scale $M$, neutrino masses from
$W$ would dominate.)
%
First, there can be $R$-symmetries  that forbid $\nu$-mass
operators in $W$ (at least at the lowest level) and allow them in
$K$. {\it E.g.}~consider the $U(1)$-symmetry with charges $q_\theta=1$
($\theta \ \equiv$ spinorial coordinate), $q_L + q_{H_2}\neq 1$. Then
$(L\cdot H_2)^2/M$ is forbidden in $W$, but $K$ can contain operators
like   those of eq.(\ref{KL}), which generate $\nu$-masses as long as
$\langle F^*_{H_1} \rangle\sim -\mu \langle H_2 \rangle\neq 0$.  This
works fine for $\kappa'$, but no $R$-symmetry can allow $\kappa$ and
forbid  $(L\cdot H_2)^2$ in $W$ if $W\supset \mu H_1H_2$.\footnote{
This can be seen by working out the charges or by noticing that
the operator $\kappa$ can be removed from $K$ by
a redefinition of the field $H_1$, and if there is a $\mu$-term, such
redefinition generates a term $(L\cdot H_2)^2$ in $W$, albeit of
magnitude automatically suppressed by  ${\cal O}(\mu/M)$.}
However the $\mu$-term can be induced through a
Giudice-Masiero mechanism \cite{GM}, from an operator
$\overline{T}H_1 H_2/M$ in the K\"ahler potential, where $T$ is a
field driving SUSY breaking, {\it i.e.}~$\langle F_T \rangle\neq 0$.
In this case $W$ contains a term $\sim \langle F_T \rangle T$  which
generates a $\mu$-term ($\mu=-\langle F_T \rangle/M$) after the
redefinition $T\rightarrow T-H_1 H_2/M$.  
Still, effective
operators like  $\sim \langle F_T \rangle (L\cdot H_2)^2\subset W$ or
$\sim \overline{T} (L\cdot H_2)^2\subset K$ can be allowed, but they induce
$\nu$-masses of the same order. 
Similar conclusions
are reached if the breaking of SUSY is communicated by gravitational
interactions, so that $m_{3/2}\sim\langle W_{hidden} \rangle /M_p^2$,
and the effective $\mu$-term is generated by an operator $\propto
H_1H_2$ in $K$. For simplicity, we do not consider
this kind of contributions to $M_\nu$. Actually, 
if one adds more fields to the MSSM it is easy to find some symmetry
that allows a $\mu$-term and the L-violating operators in $K$, but forbids
them in $W$, {\it e.g.}~if L-violation is spontaneously induced by the
vev of a field $\Phi$ (with lepton number L$=2$) with a coupling
$\overline\Phi(L_i\cdot H_2)(L_j\cdot\overline{H}_1)$ in $K$.

Second, L is an anomalous global symmetry, which is violated at 
the  non-perturbative
level. Actually, as any other global symmetry, it is expected to be
violated within any fundamental theory that includes gravity (for a
discussion of these points see {\it e.g.}~\cite{Witten}). For instance, 
in string theory the (global
symmetry) violating effects typically appear as non-perturbative
effects ({\it e.g.}~world-sheet contributions). On the other hand, if these
effects are non-perturbative in the string (-loop) expansion sense,
they are much more likely to give sizeable contributions to the
K\"ahler potential than to the superpotential (the latter is protected
by discrete symmetries) \cite {dKnp}. In general, it is clear that $W$
is much more protected than $K$ against non-perturbative (and of course
perturbative) corrections, so it makes sense to suppose that the 
L-violating operators appear in $K$ rather that in $W$.

{\bf 4. RG running.} If the scale $M$ is not too close to the electroweak
scale, the dominant effect of radiative corrections to neutrino masses and 
mixing angles is a logarithmic effect of order 
$\ln(M/M_W)$ that can be computed most easily by renormalization group
techniques: running the operators $\kappa$ and $\kappa'$ from
the scale $M$ down to $M_W$.
The renormalization group equations (RGEs) for these 
operators can be easily extracted from the one-loop correction to the
K\"ahler potential (computed for general non-renormalizable 
theories in ref.~\cite{Andrea}). They are \cite{progress} 
($16\pi^2 t\equiv \ln Q$)
\bea
\label{kappa2}
{d\kappa\over dt}&=&u\kappa+
P_E\kappa-\kappa P_E^T+2(P_E\kappa-\kappa^T P_E^T)\ ,\\
\label{kappa1}
{d\kappa'\over dt}&=&u'\kappa'-
P_E\kappa'-\kappa' P_E^T\ ,
\eea
where $u={\mathrm Tr}(3Y_U^\dagger Y_U+
3Y_D^\dagger Y_D+Y_E^\dagger Y_E)-3g^2-{g'}^2$, 
$u'=2[{\mathrm Tr}(3Y_D^\dagger Y_D+
Y_E^\dagger Y_E)+g^2+{g'}^2]$ and 
$P_E\equiv Y_EY_E^\dagger$. $Y_E$ is the matrix of leptonic Yukawa
couplings, $W\supset Y_E^{ij}H_1\cdot L_i E_{Rj}$, 
($Y_U$, $Y_D$ are those of $u$- and $d$-quarks). $g$, $g'$ are
the $SU(2)_L$, $U(1)_Y$ gauge couplings, respectively.

We see that there is no mixing between $\kappa$ and 
$\kappa'$, as was to be expected. (A supersymmetric theory
does not mix $H_2$ and $\overline{H}_1$, which make the difference between
$\kappa'$ and $\kappa$.) The RGE for $\kappa'$, 
eq.~(\ref{kappa1}), is of the standard form and similar to
the RGE for the neutrino mass matrix in the conventional scenario, 
{\it i.e.}~when the masses arise from operators in $W$ \cite{RGMSSM}:
there is a universal evolution given by $u'$
and a RG-change of the texture of $\kappa'$ induced by the $P_E$-terms. 
The stability analysis of neutrino mass splittings and mixings is similar to
that of refs.~\cite{cein,rgsusy}. In the following we will set 
$\kappa'=0$ and focus on $\kappa$. In
models with non-zero $\kappa'$ its contributions to the neutrino
masses can be simply added to those of $\kappa$, with no interference
between their runnings.

The structure of the RGE for $\kappa$ is quite remarkable. In order
to better appreciate it, it is convenient to split $\kappa$
in its symmetric and antisymmetric parts and to remember that
$\kappa_S$ contributes directly to neutrino masses, while
$\kappa_A$ does not, see eq.~(\ref{nufromK}). From eq.~(\ref{kappa2}) we 
get:
\bea
\label{kappaS}
{d\kappa_S\over dt}&=&u\kappa_S
+P_E\kappa_A-\kappa_AP_E^T\ ,\\
{d\kappa_A\over dt}&=&u\kappa_A
+P_E\kappa_S-
\kappa_S P_E^T
+2(P_E\kappa-
\kappa^T P_E^T)\ .
\label{kappaA}
\eea
Both equations contain a universal piece and a potentially
texture-changing piece, but for $\kappa_S$ this important last term 
depends only on $\kappa_A$ and {\it not} on $\kappa_S$ itself.
This peculiar behaviour is the result of a cancellation between 
(flavour-changing) wave-function-renormalization corrections 
for the leptonic legs and vertex radiative corrections (coming
from the non-minimal character of the K\"ahler potential and absent for 
couplings in $W$). This cancellation 
seems to be accidental (it does not hold at two-loops).
This peculiarity of (\ref{kappa2}) leads to the first interesting result 
concerning radiative corrections:
\begin{description}
\item [ 3)] If $\kappa_A(M)=0$ the texture of $\kappa_S$ is quite stable.
If also $\kappa'(M)=0$, neutrino mixing angles {\it do not run at one-loop}.
\end{description}
This is the only instance that we know of a model for neutrino masses with such
behaviour. If the full theory beyond $M$ is able to generate neutrino parameters
in agreement with experiment, the often dangerous radiative corrections will 
not spoil this [if the conditions listed in 3) hold].

Another interesting effect that eqs.~(\ref{kappaS},\ref{kappaA}) 
can accommodate is the
\begin{description}
\item [ 4)] {\it radiative generation of a large mixing angle}
starting from a small one, in a natural way.
\end{description}
This idea has been much investigated in the literature but, as shown
in \cite{cein,nogomax}, it requires a significant amount of
fine-tuning in order to work. This is because, whenever radiative
effects are large (as they should be to generate the large mixing) the
usual RGEs for neutrino mass matrices imply that only small-mixings
are infrared fixed points, while maximal mixings are not stable.  The
unconventional form of eq.~(\ref{kappaS}) allows us to circumvent that
difficulty. Consider  for simplicity two-flavour oscillations 
[$\nu_\mu\rightarrow\nu_\tau$ oscillations for atmospheric
neutrinos; $\nu_e\rightarrow(\cos\theta_{atm}
\nu_\mu+\sin\theta_{atm}\nu_\tau$) oscillations for solar
neutrinos]. Let us  start at $M$ with [$s, a={\cal O}(1)$]  
\be 
\kappa_S(M)=\pmatrix{s & 0\cr 0 & s}\ ,
\;\; \kappa_A(M)=\pmatrix{ 0 & a\cr -a & 0}\ ,
\ee 
{\it i.e.}~with diagonal and degenerate neutrino masses, as
given by $\kappa_S$, but with a non-zero antisymmetric matrix
$\kappa_A$. The mass matrix at low-energy is obtained by
integrating the RGEs (\ref{kappaS},\ref{kappaA}) and, in one-loop
leading-log approximation, the result is
\be 
M_\nu\propto\kappa_S(M_W)= \pmatrix{s' & a\epsilon \cr  a
\epsilon & s'}\ ,
\label{max}
\ee
with
\be 
\label{univ}
\hspace{-0.2cm}s'=s\left(1-{u\over 16\pi^2}\ln{M\over M_W}\right),\;\;\;
\epsilon={y^2\over 16\pi^2}\ln{M\over M_W}\ .  
\ee
Here $y$ is some combination of leptonic Yukawa 
couplings, dominated
by that of the $\tau$-lepton, and in the following, a prime indicates
the universal RG-scaling down to $M_W$ (as for $s'$ above). 
$M_\nu$ in (\ref{max}) has maximal mixing, which has
been naturally induced by the RG-running, as promised. 
As long as the initial masses are not exactly degenerate, the final 
mixing is not exactly maximal.

%

On the other hand, the RG-generated neutrino mass splitting is similar to
that in conventional scenarios. In particular, it should be remarked
that $\epsilon$ 
can be of the right order of
magnitude to account for the small solar mass splitting of the LMA
solution,  as was shown already in \cite{cein0}. Actually, if $\tan
\beta$ is large (and thus $y$), even the atmospheric mixing could be 
generated in this way.

{\bf 5. Textures.}
Let us now explore what can be expected for the textures of the $3\times 3$ 
neutrino mass matrix, $M_\nu$, and their phenomenological viability. If one  
starts with $\kappa'(M)=\kappa_A(M)=0$ and any texture for 
$\kappa_S(M)$, then, according to the result 3), this texture will be 
stable under radiative corrections. However, since it is possible to generate 
entries of $M_\nu$ radiatively, as explained above, it is interesting to 
examine to what extent a realistic $M_\nu$ can be produced in this way.

The simplest case is $\kappa'(M)=\kappa_S(M)=0$ [{\it i.e.},~$K$ contains 
just the $(L_i\cdot L_j)(H_2\cdot \overline{H}_1)$ operator]. 
Then $M_\nu$ arises as a purely radiative effect [thus 
a smaller $M$ (by a $\sim 10^{-2}$ factor] gives $m_\nu$'s of the right 
order). Denoting 
\be
\kappa_A(M)=\pmatrix{
 0 & a & b \cr
-a & 0 & c \cr
-b &-c & 0 }\ ,
\ee
the one-loop leading-log integration of (\ref{kappaS},\ref{kappaA}) 
gives
\be
\label{rgpiece}
M_\nu\propto\kappa_S(M_W)=
\pmatrix{
 0                  & a\epsilon_{\mu e}   & b\epsilon_{\tau e} \cr
 a \epsilon_{\mu e} &     0               & c\epsilon_{\tau\mu}\cr
 b\epsilon_{\tau e} & c\epsilon_{\tau\mu} &       0            }\ ,
\ee
where
\be
\epsilon_{\alpha\gamma}={1\over 16\pi^2}(y_\alpha^2-y_\gamma^2)\ln{M\over M_W}\ ,
\ee
and $y_\alpha\equiv\{h_e,h_\mu,h_\tau\}/\cos\beta$ are the leptonic Yukawa couplings
in the matrix $Y_E$ (which can be taken to be diagonal). This is exactly a
\begin{description}
\item [ 5)] 
(RG-generated) 'Zee'-type SUSY texture for $M_\nu$.
\end{description}
This texture has been confronted with experimental results
and good agreement requires \cite{Zeeanalysis} 
that $c\epsilon_{\tau\mu}\ll b\epsilon_{\tau e}\sim a\epsilon_{\mu e}$
(plus small deviations from the
pure Zee texture to avoid too maximal $\theta_{sol}$).
This condition
would require the hierarchy $c\ll b$ and the tuning $b/a\sim h_\mu^2/h_\tau^2$.
The origin of such hierarchies is not clear. They might result from
having different mass scales in $\kappa_A$. Alternatively, 
it is amusing to note that if $\kappa_A$ arises as a radiative effect
in the form $\kappa_A \sim A_1 P_E A_2
- (A_1 P_E A_2)^T$, where $A_{1,2}$ are generic 
antisymmetric matrices, then
$a\propto h_\tau^2$, $b\propto h_\mu^2$, $c\propto h_e^2$, as required.
In any case, the neutrino spectrum is $m_1\simeq -m_2 \gg m_3$.

The textures become more involved once $\kappa_S$ and $\kappa'$ 
are non-zero. Still, it is nice to note that one could 
generate all the mixings radiatively. {\it E.g.}~starting with $\kappa'=0$ 
(for simplicity) and  $\kappa_S={\mathrm diag}(s_1, s_2, s_3)$ it is clear 
that the low-energy mass matrix
\be
M_\nu\propto\kappa_S(M_W)=
\pmatrix{
 s'_1                  & a\epsilon_{\mu e}   & b\epsilon_{\tau e} \cr
 a \epsilon_{\mu e} &     s'_2               & c\epsilon_{\tau\mu}\cr
 b\epsilon_{\tau e} & c\epsilon_{\tau\mu} &       s'_3            }\ ,
\label{max2}  
\ee
contains the exact number of parameters necessary to fit the three
$\nu$-masses and mixing angles.   One could go further and demand
the absence of unnatural fine-tunings between the entries of 
(\ref{max2}). An interesting possibility is $s'_1\simeq s'_2 \simeq 
s'_3\sim c$ (which can be taken $\sim 1$); $b\simeq 0$.  Then
the masses are quasi-degenerate, $m_1\simeq m_2 \simeq m_3\equiv m'_0$
($\simlt 0.3$ eV  to avoid too fast $0\nu\beta\beta$-decay), and
the atmospheric angle and mass splitting  are radiatively generated if
$\tan \beta$ is large, $\sim 34\ (0.3\ {\rm eV}/m'_0)$.  However, to
avoid a solar splitting as large as the atmospheric one requires a
certain fine-tuning, $|s'_1-s'_2|\sim |c\epsilon_{\tau\mu}|$.  The solar
angle remains undetermined at this stage, but is generated once
$\kappa_S$ is perturbed with ({\it e.g.}~two-loop) corrections in all
entries. Then, it is in principle easy to accommodate the solar mixing 
and mass splitting with no further tunings. So, we conclude that
\begin{description}
\item [ 6)] To agree with experiment, textures (\ref{max2}) require some 
tuning of parameters, which is moderate in the quasi-degenerate case.

\end{description}

One can be less ambitious and try to generate just part of the
neutrino mixings and mass splittings radiatively, for example, those of the solar sector, assuming that the physics beyond the scale $M$ fixes
the atmospheric sector. Let us then start with $\kappa'=0$,
$\kappa_A\neq 0$ at the scale $M$ and
\[
\kappa_S(M)\propto M_\nu(M)=m_0\pmatrix
{1 & 0          &     0   \cr
 0 & 1+\Delta/2 & \Delta/2\cr
 0 & \Delta/2   & 1+\Delta/2}\ ,
\]
(with $m_0^2\gg \Delta m_{atm}^2$) which gives maximal atmospheric mixing 
and the right atmospheric mass splitting for
$\Delta=\Delta m_{atm}^2/{m'}_0^2$ while the solar mixing 
angle and splitting are both zero.
At low-energy, besides the universal re-scaling just mentioned, $\kappa_S$ 
gets a RG-correction of the form (\ref{rgpiece}) which drives the solar 
parameters 
to some non-zero values. If $a,b,c$ are all of the same order
of magnitude we can neglect the $\epsilon_{\mu e}$ term 
($\epsilon_{\mu e}\ll \epsilon_{\tau \mu}\simeq \epsilon_{\tau e}\equiv 
\epsilon_\tau$) 
and the only free parameters left are $b$ and $c$ (for a given $M$). 
The  generated solar mass splitting  is $\Delta m_{sol}^2\simeq 
2{m'}_0^2\sqrt{2b^2+c^2}\epsilon_\tau$ which can be inside the 
experimental range quite
naturally. The solar angle is a function of $r\equiv c/|b|$:
\be
\tan^2\theta_{sol}\simeq 1+r^2+r\sqrt{2+r^2}\ .
\ee
For $|c|\ll |b|$, $r\rightarrow 0$ and the solar angle gets exactly 
maximal. This is 
disfavoured now, but one gets values of $\tan^2\theta_{sol}$ inside the
experimental interval for $-1.0\leq r\leq-0.19$. (The atmospheric angle is not 
disturbed much by the radiative effects and stays nearly maximal while a 
non-zero
CHOOZ angle of order $\epsilon_\tau$ is generated which stays comfortably 
below the
experimental upper bound $\sin\theta_2<0.24$ \cite{concha}.)
We conclude that 
\begin{description}
\item[ 7)] $\Delta m_{sol}^2$ and $\theta_{sol}$ of LAMSW solar oscillations could
have a purely low-energy (RG) origin.
\end{description}
In such scenario, the more fundamental theory beyond $M$ should account
for the parameters that control the oscillations of atmospheric neutrinos only.
This would relax some of the demands on that theory ({\it e.g.}~on flavour
symmetries).

{\bf Acknowledgments:} I.N. thanks E. Dudas for interesting suggestions.

\end{document}